\documentclass[12pt]{article}
\def\beq{\begin{equation}}
\def\eeq{\end{equation}}
\def\ap#1#2#3 {Ann. Phys. (NY) {\bf#1} (19#2) #3}
\def\err#1#2#3 {{\it Erratum} {\bf#1} (19#2) #3}
\def\ib#1#2#3 {{\it ibid.} {\bf#1} (19#2) #3}
\def\ijmp#1#2#3 {Int. J. Mod. Phys. {\bf#1} (19#2) #3}
\def\jetp#1#2#3 {JETP Lett. {\bf#1} (19#2) #3}
\def\mpl#1#2#3 {Mod. Phys. Lett. {\bf#1} (19#2) #3}
\def\np#1#2#3 {Nucl. Phys. {\bf#1} (19#2) #3}
\def\pl#1#2#3 {Phys. Lett. {\bf#1} (19#2) #3}
\def\prep#1#2#3 {Phys. Rep. {\bf#1} (19#2) #3}
\def\prev#1#2#3 {Phys. Rev. {\bf#1} (19#2) #3}
\def\prl#1#2#3 {Phys. Rev. Lett. {\bf#1} (19#2) #3}
\def\sjnp#1#2#3 {Sov. J. Nucl. Phys. {\bf#1} (19#2) #3}
\def\spj#1#2#3 {Sov. Phys. JETP {\bf#1} (19#2) #3}
\def\spu#1#2#3 {Sov. Phys. Usp. {\bf#1} (19#2) #3}
\def\zp#1#2#3 {Zeit. Phys. {\bf#1} (19#2) #3}

\begin{document}
\begin{titlepage}
\begin{center}
{\Large \bf Theoretical Physics Institute \\
University of Minnesota \\}  \end{center}
\vspace{0.2in}
\begin{flushright}
TPI-MINN-01/27-T \\
UMN-TH-2011-01 \\
June 2001 \\
\end{flushright}
\vspace{0.3in}
\begin{center}
{\Large \bf  Non-factorization effects in heavy mesons and
determination of $|V_{ub}|$ from inclusive semileptonic $B$ decays.
\\}
\vspace{0.2in}
{\bf M.B. Voloshin  \\ }
Theoretical Physics Institute, University of Minnesota, Minneapolis,
MN
55455 \\ and \\
Institute of Theoretical and Experimental Physics, Moscow, 117259
\\[0.2in]
\end{center}

\begin{abstract}
The effects of spectator light quark in decays of heavy mesons are
considered, which vanish in the limit of factorization of matrix
elements of four-quark operators over the mesons. These effects include
the difference of the total widths as well as of the semileptonic decay
rates between the $D^0$ and $D_s$ mesons and also a contribution to
inclusive semileptonic decay rates of $B^0$ and $B^\pm$ into the channel
$X_u \, \ell \, \nu$ related to determination of the weak mixing
parameter $V_{ub}$. If the observed difference of the lifetimes between
$D_s$ and $D^0$ mesons is attributed to non-factorizable terms, such
terms can naturally give rise to a significant difference in inclusive
semileptonic decay rates of these mesons, and to a light-flavor
dependent contribution to decays $B \to X_u \, \ell \, \nu$. The latter
contribution affects mostly the upper part of the inclusive spectrum of
the invariant mass of the lepton pair, and may significantly exceed the
previously claimed in the literature theoretical uncertainty in
determination of $|V_{ub}|^2$ from that part of the spectrum.
\end{abstract}

\end{titlepage}

\section{Introduction}
The well known difficulty of determination of the weak mixing parameter
$|V_{ub}|$ from the inclusive decay $B \to X_u \, \ell \, \nu$ is in
discriminating this process from the much more abundant decays $B \to
X_c \, \ell \, \nu$ (see e.g. in the review in Ref. \cite{pdg} and
references therein). The experimental method used so far is to impose a
lower cutoff on the charged lepton energy at or above the kinematical
boundary for the processes with charmed hadrons in the final state.
Theoretically however a calculation of the lepton energy spectrum in the
endpoint region is prone to a substantial uncertainty \cite{pdg}, which
is unlikely to be resolved without knowledge of $B$ meson structure
functions \cite{twist}.

A somewhat more theoretically tractable scheme arises \cite{bll} if a
cutoff is applied to the invariant mass squared $q^2$ of the lepton
pair, $q^2 \ge q_0^2$, instead of the single-lepton energy. Clearly, the
condition $q^2_0 \ge (M_B-M_D)^2$ makes a sample of semileptonic decay
events free from the background from $B \to X_c \, \ell \, \nu$. The
total decay rate into the part of the phase space limited by such
constraint can be analysed, including nonperturbative contributions as
well as the perturbative one, by applying the operator product expansion
(OPE) for inclusive decay rates \cite{sv1,sv2}. Most recently an
analysis along these lines was performed in Ref.\cite{bn}.

The difficulty arising in this approach is that the ``short distance"
parameter in the OPE for the constrained inclusive decay rate is
determined \cite{bn} by the typical momentum $\mu_c$ of the $u$ quark in
the constrained kinematical region, where its maximal value is $(M_B^2 -
q_0^2)/2M_B$, which becomes slightly smaller than the mass of the
charmed quark $m_c$, even if the cutoff parameter is chosen at exactly
the kinematical limit for charmed final states, $q^2_0 \ge (M_B-M_D)^2$.
This puts the theoretical status of the OPE for the constrained
inclusive rate on par with or even somewhat worse than that of an OPE
based analysis of total inclusive weak decay rates of charmed mesons and
baryons. It is well known that in the latter analysis the
nonperturbative terms in OPE, whose relative contribution is suppressed
as $m_c^{-2}$ and $m_c^{-3}$, are essentially as significant as the
``leading" perturbative (parton) term, which behavior prominently
manifests itself in the large differences of the lifetimes both between
the charmed mesons and baryons and within the meson and baryon
light-flavor multiplets. This simple observation illustrates the acute
necessity of considering the higher terms, formally suppressed by
$\mu_c^{-2}$ and $\mu_c^{-3}$, in the OPE based analysis of the
constrained decay rate of $B \to X_u \, \ell \, \nu$. The leading
perturbative term as well as the nonperturbative contribution
proportional to $\mu_c^{-2}$ were considered in Ref.\cite{bn}, while the
contribution relatively suppressed by $\mu_c^{-3}$ was relegated to a
theoretical uncertainty. The magnitude of the uncertainty, arising from
this source, was estimated (essentially by guessing) to correspond to
about 5\% theoretical error in $|V_{ub}|$.
The purpose of this paper is to discuss the contribution of the third
term in OPE which arises from non-factorizable terms in matrix elements
of four-quark operators over heavy mesons, and to analyse, to an extent,
the numerical significance of this contribution. The impact of
non-factorizable terms on semileptonic decays of
$D$ and $B$ mesons as well as on the difference of lifetimes of $D_s$
and $D^0$ has been discussed previously \cite{bu1,bu2,nu}. It is however
worthwhile, in view of the recent discussion \cite{bll,bn} of
theoretical uncertainty in determination of $|V_{ub}|$, to point out the
effects of the non-factorizable terms in this particular issue, and to
present specific estimates relevant to probing these terms from data on
charmed mesons. It is emphasized here that these terms, generally
different for $B^0$ and $B^\pm$ mesons (as first pointed out in
Ref.\cite{bu1}), can be quite
essential in the constrained inclusive semileptonic decay rates due to
the $b \to u$ transition, if the magnitude of these effects is
estimated, following the arguments of Ref.\cite{bu2},
from the observed difference of lifetimes of the $D_s$ and $D^0$ mesons.
A more direct evaluation of the light-flavor
dependent part of the relevant non-factorizable terms would be possible
if the difference of the total semileptonic decay rates between $D_s$
and $D^0$ mesons could be measured experimentally.

As will be discussed in Sec. 2 this contribution is described, as usual,
by dimension 6 four-quark operators. The matrix element of the relevant
combination of the operators over the $B$ mesons vanishes in the limit
of factorization. However, also as usual, the considered term is greatly
enhanced numerically, so that a violation of the factorization relation
by 10\% makes this contribution about twice larger than the estimate of
the corresponding uncertainty in Ref.\cite{bn}. Furthermore, the effect
explicitly depends on the light quark flavor and thus is generally
different in the decays of $B^0$ and $B^\pm$ mesons.
In Sec. 3 it is argued that a deviation from factorization of such
magnitude would not be unusual, and in fact is possibly indicated by the
observed difference of the lifetimes of the $D_s$ and $D^0$ mesons, if
that difference is analysed within the same OPE framework. Moreover, the
discussed effect in the constrained decay rate of $B \to X_u \, \ell \,
\nu$ is very closely related to a similar effect in semileptonic charm
decays, in particular to the yet unmeasured difference of total
inclusive {\it semileptonic} decay rates of $D_s$ and $D^0$. Thus a
quantitative understanding of the discussed contribution would become
possible, once this difference of the semileptonic decay rates is
measured experimentally. The concluding Section 4 contains discussion
and summary.

\section{The third term in OPE for constrained inclusive decay rate.}
The optical theorem of the scattering theory relates the total decay
rate $\Gamma_H$ of a hadron $H_Q$ containing a heavy quark $Q$ to the
imaginary part of the `forward scattering amplitude'. For the case of
inclusive decays generated by a particular term $L_W$ in the weak
interaction Lagrangian  the latter amplitude is described by the
following effective operator
\beq
L_{eff}=2 \,{\rm Im} \, \left [ i \int d^4x \, e^{ipx} \, T \left \{
L_W^\dagger(x),
L_W(0) \right \} \right ]~,
\label{leff}
\eeq
in terms of which operator (at $p^2=m_Q^2$) the total decay rate is
given by\footnote{
The non-relativistic normalization
for the {\it heavy} quark states is used here: $\langle
Q | Q^\dagger Q | Q \rangle =1$.}
\beq
\Gamma_H=\langle H_Q | \, L_{eff} \, | H_Q \rangle~.
\label{lgam}
\eeq
Using in eq.(\ref{leff}) the term
\beq
L_{ub}={G_F \, V_{ub} \over \sqrt{2}} \, (\bar u \, \gamma_\mu \,
(1-\gamma_5) \, b) \, \ell_\mu
\label{lub}
\eeq
with $\ell_\mu=(\bar \ell \, \gamma_\mu \, (1-\gamma_5) \, \nu)$ in
place of $L_W$, one would find the total inclusive decay rate of $B \to
X_u \, \ell \, \nu$. The effective operator (\ref{leff}) is evaluated
using short-distance OPE. The leading term in the expansion describes
the perturbative decay rate, while subsequent terms containing operators
of higher dimension describe the nonperturbative contributions.
In particular, the leading term with no QCD radiative corrections
applied, gives for the total inclusive rate of the decays $B \to X_u \,
\ell \, \nu$ the familiar expression
\beq
\Gamma_0 = { G_F^2 \, |V_{ub}|^2 \, m_b^5 \over 192 \, \pi^3}~.
\label{g0}
\eeq

For a constrained inclusive rate in the case of semileptonic decays and
with selection in the invariant mass of the lepton pair, one should
rather consider, instead of the full correlator (\ref{leff}), a
correlator of the quark currents, and then perform integration over the
lepton pair phase space with the weight function corresponding to the
considered selection of events \cite{bu1}. However for the third term in
the OPE, suppressed by $m_Q^{-3}$ and expressed through four-quark
operators of dimension 6, this procedure is not really necessary, since
{\it formally} this contribution corresponds to an infinitesimally small
momentum carried away by the final light quark, and all the momentum of
the initial heavy quark flowing through the lepton pair. Thus this
contribution {\it formally} comes only from $q^2=m_Q^2$, and does not
depend on the lower cutoff $q_0^2$. (This behavior  of the discussed
nonperturbative effect, first pointed ou in Ref.\cite{bu1}, was also
used in the estimate
of Ref.\cite{bn}.) Therefore the effect of this term constitutes a fixed
fraction of the total rate of the decays $B \to X_u \, \ell \, \nu$.

It is quite straightforward to find the expression for the third term in
the expansion of the correlator (\ref{leff}) appropriate for the decays
$B \to X_u \, \ell \, \nu$ \cite{sv2,bu1,bu2,sv3,mv} in terms of
four-quark
operators normalized at $\mu=m_b$:
\beq
L^{(3)}_{b \to u \ell \nu}=-{2 \, G_F^2 \, |V_{ub}|^2 \, m_b^2 \over 3
\, \pi} \, \left ( O^u_{V-A}-O^u_{S-P} \right )~,
\label{l3bu}
\eeq
where the following notation \cite{ns} is used for the relevant
four-quark operators:
\begin{eqnarray}
&&O^q_{V-A}=(\bar b_L \gamma_\mu q_L)(\bar q_L \gamma_\mu b_L)\,,
~~O^q_{S-P}=(\bar b_R  q_L)(\bar q_L  b_R)\,, \nonumber \\
&&T^q_{V-A}=(\bar b_L t^a \gamma_\mu q_L)(\bar q_L t^a \gamma_\mu
b_L)\,, ~~T^q_{S-P}=(\bar b_R t^a q_L)(\bar q_L t^a b_R)\,.
\label{ops}
\end{eqnarray}
(The operators $T$, containing the color generators $t^a$, will appear
in further discussion.)

The matrix elements of the four-quark operators over the $B$ mesons are
parameterized \cite{ns} as
\beq
\langle B | O^u_{V-A} | B \rangle = {f_B^2 \, m_B \over 8} \, B_1\, , ~~
\langle B | O^u_{S-P} | B \rangle = {f_B^2 \, m_B \over 8} \, B_2\,,
\label{param}
\eeq
where $f_B$ is the meson annihilation constant, and $B_1$ and $B_2$ are
generally phenomenological parameters, ``bag constants". In the limit of
naive factorization, i.e. where the product of bilinear operators is
saturated by only the vacuum insertion, these parameters are $B_1=B_2=1$
for the $B$ meson containing the same light quark as the operator
($B^\pm$ mesons for the case of the operators $O^u$), and are zero for
the other two mesons ($B_d$ and $B_s$ in this case). Clearly, in this
limit the matrix element of the four-quark operator in eq.(\ref{l3bu})
over either of the $B$ mesons is vanishing\footnote{The naive
factorization for the discussed effect actually correspond to the
annihilation decay $B^\pm \to \ell \nu$, which obviously is forbidden by
chirality in the limit where the lepton mass is neglected.}.  However,
as useful as the factorization relation might be for other estimates,
one does not expect this relation to be valid with better than about
10\% accuracy, and it is conventionally
implied that deviations of such order of magnitude are present. (For
numerical estimates of non-factorizable terms see e.g.
\cite{vc,blls,pu}.)
Moreover, it should be emphasized that these deviations refer not only
to the relation between the constants: $B_1 =B_2$ for the meson with the
same flavor as that of the quark $q$ in the operator $O^q$, but that
these constants are generally non zero when the light quark flavors do
not match \cite{bu2}. In other words, there are separate sets of the
constants $B$
describing the matrix elements $B_{q' q}=\langle B_{q'}|O^q| B_{q'}
\rangle$ for $q'=q$ and for $q' \neq q$. Taking into account also the
violation of the flavor SU(3) would further proliferate the constants
$B$.

In any case, a violation of the relation $B_1 = B_2$ by about 10\% would
not be surprising at all.
On the other hand, a deviation from the exact factorization of such
moderate magnitude would be quite significant for the discussed effect
in the $B \to X_u \, \ell \, \nu$ decays. Indeed, in terms of the ``bag
constants" the contribution $\delta \Gamma$ to the decay rate  of the
term (\ref{l3bu}) is expressed according to equations (\ref{lgam}) and
(\ref{param}) as
\beq
\delta \Gamma \approx { G_F^2 \, |V_{ub}|^2 \,f_B^2 \, m_b^3 \over 12 \,
\pi} \left (B_2-B_1 \right )~.
\label{dgam}
\eeq
Comparing this with the dominant semileptonic decay rate of $B \to X_c
\, \ell \, \nu$, and taking into account the known semileptonic
branching ratio $B_{sl}(B)$, one estimates that the discussed
nonperturbative contribution corresponds
in terms of the branching ratio to the correction
\beq
\delta B(B \to X_u \, \ell \, \nu) \approx 3.9 \, \left ({f_B \over 0.2
\, GeV } \right )^2 \, \left ( {B_2 - B_1 \over 0.1} \right ) \,
|V_{ub}|^2~,
\label{dbr}
\eeq
which easily can exceed by a substantial amount the previous estimate
\cite{bn} of this correction as $2.0 \, |V_{ub}|^2$ if the deviation
from factorization is, as expected, at a 10\% level.

\section{Non-factorization effects in $D$ mesons}
The deviation from naive factorization in mesons can in fact be probed
more quantitatively from the data on charmed $D$ mesons. Indeed, in the
limit of factorization, the OPE for the dominant Cabibbo unsuppressed
nonleptonic decays due to $c \to s \, u \, \bar d$ predicts equal rates
of decay for $D_s$ and $D^0$ mesons \cite{sv2,sv3}. Experimentally it is
known by now \cite{pdg} that there is a noticeable difference in
lifetimes of these mesons: $\tau(D_s)/\tau(D^0) = 1.20 \pm 0.025$, which
cannot be described by spectator dependent effects in Cabibbo suppressed
decay channels, or by the flavor SU(3) breaking \cite{bu2}. Although
this discrepancy can be merely attributed to
the overall inaccuracy of the OPE in the inverse of the charmed quark
mass\footnote{In this respect the situation is no better for the
expansion of a constrained inclusive rate of the decays $B \to X_u \,
\ell \, \nu$, where the expansion is governed by $\mu_c < m_c$
\cite{bn}.}, a more constructive approach would be to attempt describing
this difference in lifetimes as due to deviations from factorization
(see also in \cite{bu2,nu}).  As
will be discussed further in this section such approach also can be
tested by measuring the difference of inclusive {\it semileptonic} decay
rates of the $D_s$ and $D^0$ mesons, which difference is somewhat more
directly related to the discussed matrix elements over the mesons in
equations (\ref{dgam}) and (\ref{dbr}).

In the limit of flavor SU(3) symmetry the difference of the dominant
inclusive nonleptonic decay rates of $D^0$ and $D_s$ mesons can be
written \cite{sv3} in terms of matrix elements of four-quark operators
(normalized at $\mu=m_c$) as
\beq
\Gamma(D^0)-\Gamma(D_s)={2 \, G_F^2 \, \cos^4 \theta_c \, m_c^2 \, f_D^2
\, m_D \over 9 \pi} \, C_+ \, C_- \, \left ( B_1^{ns} - B_2^{ns} - {3
\over 4} \, \varepsilon_1^{ns} \, + {3 \over 4} \, \varepsilon_2^{ns}
\right )~,
\label{dgamd}
\eeq
where $\theta_c$ is the Cabibbo angle, $C_+$ and $C_-$ are the well
known short-distance QCD renormalization coefficients for nonleptonic
weak interaction: $C_-=C_+^{-2}=(\alpha_s(m_c)/\alpha_s(m_W))^{12/25}$,
and the flavor non-singlet coefficients $B$ and $\varepsilon$
parameterize the following differences of the matrix elements:
\begin{eqnarray}
&&{1 \over 2} \, \langle O^s_{V-A} - O^u_{V-A} \rangle_{D_s-D^0}= {f_D^2
\, m_D \over 8} \, B^{ns}_1~, \nonumber \\
&&{1 \over 2} \, \langle O^s_{S-P} - O^u_{S-P} \rangle_{D_s-D^0}= {f_D^2
\, m_D \over 8} \, B^{ns}_2~, \nonumber \\
&&{1 \over 2} \, \langle T^s_{V-A} - T^u_{V-A} \rangle_{D_s-D^0}= {f_D^2
\, m_D \over 8} \, \varepsilon^{ns}_1~, \nonumber \\
&&{1 \over 2} \, \langle T^s_{S-P} - T^u_{S-P} \rangle_{D_s-D^0}= {f_D^2
\, m_D \over 8} \, \varepsilon^{ns}_2~,
\label{nsparam}
\end{eqnarray}
where the operators $O$ and $T$ are the same as in eq.(\ref{ops}) with
the $b$ quark being replaced by $c$ and the notation $\langle X
\rangle_{A-B} \equiv \langle A | X | A \rangle - \langle B | X | B
\rangle$ is used. (The parameters $\varepsilon_1$ and $\varepsilon_2$
both vanish in the limit of factorization.) It should be also mentioned
that no attempt is being made here to allow for the breaking of the
flavor SU(3) symmetry, thus no distinction is made between the
annihilation constants or masses of the $D_s$ and $D_0$ mesons.

The expression (\ref{dgamd}) for the difference of the total decay rates
corresponds numerically to
\beq
\Gamma(D^0)-\Gamma(D_s) \approx 3.3 \,  \left ({ f_D \over 0.22 \, GeV}
\right )^2 \,  \left ( B_1^{ns} - B_2^{ns} - {3 \over 4} \,
\varepsilon_1^{ns} \, + {3 \over 4} \, \varepsilon_2^{ns} \right ) \,
ps^{-1}~.
\label{dgamdn}
\eeq
Comparing this estimate with the experimental value for the difference
of the total decay rates: $0.41 \pm 0.05 \, ps^{-1}$, one arrives at an
estimate of corresponding combination of the non-singlet factorization
parameters:
\beq
B_1^{ns} - B_2^{ns} - {3 \over 4} \, \varepsilon_1^{ns} \, + {3 \over 4}
\, \varepsilon_2^{ns} \approx 0.12~,
\label{estp}
\eeq
which perfectly complies with the understanding that non-factorizable
contributions are at a level of about 10\%.

The estimate (\ref{estp}) of the non-factorizable terms however can
serve only as a semi-quantitative indicator of the magnitude of the
spectator effects in the inclusive rate of the processes $B \to X_u \,
\ell \, \nu$ described by a different combination of the factorization
parameters in eq.(\ref{dbr}) than in eq.(\ref{estp}). A somewhat more
direct test of the relevant combination of the parameters would be
possible from the difference of the total semileptonic decay rates of
$D_s$ and $D^0$ mesons. Indeed, in the limit of flavor SU(3) symmetry
this difference arises only in the decays due to $c \to s \, \ell \,
\nu$ and is given in terms of the operators normalized at $\mu=m_c$ as
\begin{eqnarray}
&&\Gamma_{sl}(D^0)-\Gamma_{sl}(D_s)={G_F^2 \, \cos^2 \theta_c \, m_c^2
\, f_D^2 \, m_D \over 12 \pi} \, \left (  B_1^{ns} - B_2^{ns} \right )
\nonumber \\
&&\approx 1.1  \, \left ({ f_D \over 0.22 \, GeV} \right )^2 \,  \left (
B_1^{ns} - B_2^{ns}  \right ) \, ps^{-1}~.
\label{dgsld}
\end{eqnarray}
Given that the total semileptonic decay rate of the $D^0$ meson is
approximately $0.16 \, ps^{-1}$ \cite{pdg}, the discussed difference can
easily amount to a quite sizeable fraction of the semileptonic rate,
provided that $ B_1^{ns} - B_2^{ns} \sim 0.1$. A more precise
calculation of the relative effect of this difference in terms of the
parameters $B_1^{ns}$ and $B_2^{ns}$ is somewhat difficult to achieve at
present. The problem here is that the observed semileptonic decay rate
of $D^0$ already includes, in comparison with the `bare' parton rate,
quite substantial negative QCD radiative corrections as well as the
$O(m_c^{-2})$ corrections \cite{buv}, while neither of these is included
in eq.(\ref{dgsld}). If compared with the `bare' parton semileptonic
decay rate of $c \to s \, \ell \, \nu$, the discussed effect is of a
somewhat more moderate relative magnitude:
\beq
{\Gamma_{sl}(D^0)-\Gamma_{sl}(D_s) \over \Gamma_0(c \to s \, \ell \,
\nu)} = 3.4 \, \left ({ f_D \over 0.22 \, GeV} \right )^2 \,  \left (
B_1^{ns} - B_2^{ns}  \right )~.
\eeq
In either case however the difference should be quite conspicuous and
can amount to several tens percent, provided that $|B_1^{ns} - B_2^{ns}|
\sim 0.1$.

A measurement of the difference of the inclusive semileptonic decay
rates of the $D^0$ and $D_s$ mesons would make it possible to more
reliably predict the difference of the corresponding decay rates between
$B^0$ and $B^\pm$ mesons: $\Gamma(B^0 \to X_u \, \ell \,
\nu)-\Gamma(B^\pm \to X_u \, \ell \, \nu)$, which, according to the
previous discussion, is dominantly concentrated in the upper part of the
spectrum of the invariant mass of the lepton pair \cite{bu1,bu2}. At the
level of
accuracy of the present discussion the only difference between the
theoretical expressions for $B$ and for $D$ mesons arises through a
different normalization point of the four-quark operators in the
equations (\ref{dgam}) and (\ref{dgsld}). Taking into account the
`hybrid' evolution of the operators containing $b$ quark down to $\mu =
m_c$ gives the relation between the non-singlet factorization constants:
\beq
B^{ns}_1(m_b)-B^{ns}_2(m_b) = {8 \, \kappa^{1/2}+1 \over 9} \left [
B^{ns}_1(m_c)-B^{ns}_2(m_c \right ] - { 2 \, (\kappa^{1/2}-1) \over 3}
\, \left [ \varepsilon^{ns}_1(m_c)-\varepsilon^{ns}_2(m_c \right ]~,
\label{bcrel}
\eeq
where $\kappa=(\alpha_s(m_c)/\alpha_s(m_b))$. However, modulo the
unlikely case that the difference of the constants $\varepsilon$ in this
relation is much bigger than the difference between the constants $B$,
the renormalization effect is quite small, and most likely is at the
level of other uncertainties in the considered approach (such as the
accuracy of the flavor SU(3) symmetry, higher QCD corrections,
contribution of higher terms in $m_c^{-1}$, etc.). Thus with certain
reservations, one can use the approximate relation
$B^{ns}_1(m_b)-B^{ns}_2(m_b) \approx B^{ns}_1(m_c)-B^{ns}_2(m_c)$ to
relate directly the differences in the inclusive semileptonic decay
rates:
\beq
\Gamma(B^0 \to X_u \, \ell \, \nu)-\Gamma(B^\pm \to X_u \, \ell \, \nu)
\approx {|V_{ub}|^2 / |V_{cs}|^2 } \, {f_B^2 \over f_d^2} \, {m_b^3
\over m_c^3} \, \left ( \Gamma_{sl}(D^0)-\Gamma_{sl}(D_s) \right )~.
\label{apprel}
\eeq

\section{Discussion and summary}
Naturally, the largest uncertainty of the approach pursued in the
present paper is perceived to be coming from applying the operator
product expansion for the amplitude in eq.(\ref{leff}) in Minkowski
space at energy equal to the mass of the charmed quark. This raises both
the issue of the contribution of higher terms as well as that of the
applicability of quark-hadron duality at such energy. The situation is
further complicated by the poor knowledge of the hadronic matrix
elements of four-quark operators, including the deviation from the naive
factorization in mesons. Same uncertainties are pertinent in the same,
or greater, measure to a calculation of the constrained inclusive rate
of the decays $B \to X_u \, \ell \, \nu$ at $q^2 \ge (M_B-M_D)^2$. At
present these issues are unlikely to be resolved purely theoretically,
and an experimental input is acutely needed. The accuracy of the
approach can be probed by testing at a quantitative level its
theoretical predictions, most notably that of a large enhancement
\cite{mv} of the semileptonic decay rates of the strange charmed
hyperons $\Xi_c^+, \Xi_c^0$ as compared to that of $\Lambda_c$. It was
argued previously \cite{bu2,nu} that the non-factorizable terms give a
sizeable contribution to the overall inclusive semileptonic decay rate
of $D$ mesons. However, given the uncertainty in other parameters (the
charmed quark mass, higher QCD corrections, etc.) it could be somewhat
ambiguous to assess the light-flavor singlet effect of these terms in
the decay rates. As discussed in this paper, an experimental measurement
of the difference of inclusive semileptonic decay rates between $D^0$
and $D_s$, determined by the non-singlet part of the non-factorizable
terms, is more likely to shed light on the problem of the deviations
from factorization in heavy mesons. Unless a better understanding of the
non-factorizable terms is achieved,  they will stand in the way of
determining the mixing parameter $|V_{ub}|^2$ from inclusive
semileptonic decay spectra with accuracy better than $O(15\%)$,
including the method discussed in Refs.\cite{bll,bn}.

In summary. It is pointed out that non-factorizable terms can provide a
quite noticeable contribution to inclusive rates of the decays $B \to
X_u \, \ell \, \nu$, which depends on the flavor of the light spectator
quark, and is thus different for $B^0$ and $B^\pm$ mesons. This yet
unknown contribution can limit the theoretical accuracy of determining
$|V_{ub}|^2$ from the lepton spectra beyond the kinematical limit for
decays with charm in the final state. The effect of such contribution
can be substantially larger than the previously estimated \cite{bn}
uncertainty, if the non-factorizable terms are evaluated from the
experimentally observed difference of the lifetimes of $D_s$ and $D_0$
mesons. A more direct conclusion about the non-factorization effects in
charmless semileptonic decays of the $B$ mesons can be drawn once the
difference of the total semileptonic decay rates of $D^0$ and $D_s$ is
measured experimentally. If the current understanding is correct and the
non-factorizable terms amount to about 10\%, the latter difference can
amount to a significant fraction of the total semileptonic decay rate of
charmed mesons.

\section{Acknowledgements}
This work is supported in part by DOE under the grant number
DE-FG02-94ER40823.


\begin{thebibliography}{99}
\bibitem{pdg}
Particle Data Group, Eur.Phys.J. {\bf C 15} (2000) 1.
\bibitem{twist}
I.I. Bigi, M.A. Shifman, N.G. Uraltsev, and A.I. Vainshtein, Int. J.
Mod. Phys. A {\bf 9} (1994) 2467; \pl{B 328}{94}{431}. \\
M. Neubert, \prev{D49}{94}{3392}.\\
T. Mannel and M. Neubert, \prev{D50}{94}{431}.
\bibitem{bll}
C.W. Bauer, Z. Ligeti, and M. Luke, Phys.Lett. {\bf B 479} (2000) 395.
\bibitem{sv1}
M.A. Shifman and M.B. Voloshin, (1981) unpublished, presented in the
review V.A.Khoze and M.A. Shifman, \spu{26}{83}{387}.
\bibitem{sv2}
M.A. Shifman and M.B. Voloshin, \sjnp{41}{85}{120}.
\bibitem{bn}
T. Becher and M. Neubert, Cornell report CLNS 01/1737;\
[hep-ph/0105217].
\bibitem{bu1}
I.I. Bigi and N.G. Uraltsev, \np{B423}{94}{33}.
\bibitem{bu2}
.I. Bigi and N.G. Uraltsev, \zp{C62}{94}{623}.
\bibitem{nu}
N.G. Uraltsev, \ijmp{A14}{99}{4641}.
\bibitem{sv3}
M.A. Shifman and M.B. Voloshin, \spj{64}{86}{698}.
\bibitem{mv}
M.B. Voloshin, \pl{B385}{96}{369}.
\bibitem{ns}
M. Neubert and  C.T. Sachrajda, \np{B483}{97}{339}.
\bibitem{vc}
V. Chernyak, \np{B457}{95}{96}.
\bibitem{blls}
M.S. Baek, J. Lee, C. Liu, and H.S. Song, \prev{D57}{98}{4091}.
\bibitem{pu}
D. Pirjol and N. Uraltsev, \prev{D59}{99}034012.
\bibitem{buv}
I.I. Bigi, N.G. Uraltsev, and A.I. Vainshtein, \pl{B293}{92}{430}; \ [E:
{\bf B297} (1993) 477].

\end{thebibliography}
\end{document}